\documentclass[compsoc,conference,a4paper,10pt,times]{IEEEtran}
\IEEEoverridecommandlockouts
\usepackage{cite}
\usepackage{amsmath,amssymb,amsfonts}
\usepackage[T1]{fontenc}  
\usepackage{algorithmic}
\usepackage{graphicx}
\usepackage{textcomp}
\usepackage{bmpsize}
\usepackage{xcolor}
\usepackage{lipsum}
\usepackage{blindtext}
\usepackage{subfiles}
\usepackage{placeins}
\usepackage{subcaption}
\usepackage{multirow}

\usepackage{subcaption}
\usepackage{glossaries}
\usepackage[colorlinks=true,urlcolor=black]{hyperref}

\def\BibTeX{{\rm B\kern-.05em{\sc i\kern-.025em b}\kern-.08em
    T\kern-.1667em\lower.7ex\hbox{E}\kern-.125emX}}
    
\newcommand{\shorten}[1]{}

\newglossaryentry{A_B^C}{type=main,name={\ensuremath{A_B^C}},sort=A,
description={A dummy symbol }}

\newacronym{os}{OS}{Operating System}
\newacronym{cli}{CLI}{Command Line Interface}
\newacronym{api}{API}{Application Programming Interface}
\newacronym{rest}{REST}{Representational state transfer} 
\newacronym{nvd}{NVD}{National Vulnerability Database}
\newacronym{cve}{CVE}{Common Vulnerabilities and Exposures}
\newacronym{cvss}{CVSS}{Common Vulnerability Scoring System}
\newacronym{cpe}{CPE}{Common Platform Enumeration}
\newacronym{cwe}{CWE}{Common Weakness Enumeration}
\newacronym{lxc}{LXC}{Linux Containers}
\newacronym{cpu}{CPU}{Central Processing Unit}
\newacronym{ram}{RAM}{Random-Access Memory}
\newacronym{it}{IT}{Information Technology}
\newacronym{vm}{VM}{Virtual Machine}
\newacronym{cnas}{CNAs}{CVE Numbering Authorities}
\newacronym{html}{HTML}{HyperText Markup Language}
\newacronym{csv}{CSV}{Comma Seperated Values}
\newacronym{sql}{SQL}{Structured Query Language}
\newacronym{rdbms}{RDBMS}{Relational Database Management System}
\newacronym{gui}{GUI}{Graphical User Interface}
\newacronym{rhsa}{RHSA}{Red Hat Security Advisory}
\newacronym{CI/CD}{CI/CD}{Continuous Integration/Continuous Delivery}
\newacronym{er}{ER}{Entity-Relationship}
\newacronym{pcc}{PCC}{Pearson correlation coefficient}
\newacronym{kde}{KDE}{kernel density estimation}

\makeglossaries

\begin{document}

\title{Vulnerability Analysis of 2500 Docker Hub Images}

\author{ 
\IEEEauthorblockN{Katrine Wist}
\IEEEauthorblockA{\textit{Dep. of Inf. Sec. and Comm. Techn.} \\
\textit{Norwegian University of Science}\\
 and Technology (NTNU), Norway \\
katrinew0702@gmail.com}
\and
\IEEEauthorblockN{Malene Helsem}
\IEEEauthorblockA{\textit{Dep. of Inf. Sec. and Comm. Techn.} \\
\textit{Norwegian University of Science}\\
 and Technology (NTNU), Norway \\
malenehlsm@gmail.com}
\and
\IEEEauthorblockN{Danilo Gligoroski}
\IEEEauthorblockA{\textit{Dep. of Inf. Sec. and Comm. Techn.} \\
\textit{Norwegian University of Science}\\
 and Technology (NTNU), Norway \\
danilog@ntnu.no}
}

\maketitle

\begin{abstract}
The use of container technology has skyrocketed during the last few years, with Docker as the leading container platform. Docker's online repository for publicly available container images, called Docker Hub, hosts over 3.5 million images at the time of writing, making it the world's largest community of container images. We perform an extensive vulnerability analysis of 2500 Docker images. It is of particular interest to perform this type of analysis because the vulnerability landscape is a rapidly changing category, the vulnerability scanners are constantly developed and updated, new vulnerabilities are discovered, and the volume of images on Docker Hub is increasing every day. Our main findings reveal that (1) the number of newly introduced vulnerabilities on Docker Hub is rapidly increasing; (2) certified images are the most vulnerable; (3) official images are the least vulnerable; (4) there is no correlation between the number of vulnerabilities and image features (i.e., number of pulls, number of stars, and days since the last update); (5) the most severe vulnerabilities originate from two of the most popular scripting languages, JavaScript and Python; and (6) Python 2.x packages and jackson-databind packages contain the highest number of severe vulnerabilities. We perceive our study as the most extensive vulnerability analysis published in the open literature in the last couple of years. 

\end{abstract}


\section{Introduction}
\vspace{-0.25cm}
Container technology has been known for a long time in Linux systems through Linux Containers (LXC), but it was not commonly used until a decade ago. The introduction of Docker in \cite{avram2013docker} made the popularity of containerization rise exponentially. Container technology has revolutionized how software is developed and is seen as a paradigm shift. More concretely, containerization is considered as a beneficial technique for Continuous Integration/Continuous Delivery (CI/CD) pipelines; it is providing an effective way of organizing microservices; it is making it easy to move an application between different environments; and in general, it is simplifying the whole system development life cycle. 

Software containers got its name from the shipping industry since the concepts are fundamentally the same. A software container is code wrapped up with all its dependencies so that the code can run reliably and seamlessly in any computer environment isolated from other processes. 
Hence, containers are convenient, lightweight, and fast technology to achieve isolation, portability, and scalability.

Container technology is replacing virtual machines continuously, and the trend is that more companies are choosing to containerize their applications. Gartner predicts that more than 70\% of global companies will have more than two containerized applications in production by 2023. This is an increase from less than 20\% in 2019.\footnote{Gartner: 3 Critical Mistakes That I\&O Leaders Must Avoid With Containers} With the advent of 5G communication technology, it seems that container technology and particularly Docker is finding new venue for application in the domain of network slicing, network management, orchestration and in 5G testbeds \cite{esmaeily2020cloud}.
\shorten{
There are alternatives to Docker, to mention a few: LXC\footnote{https://linuxcontainers.org/}, rkt\footnote{https://coreos.com/rkt/}, Apache Mesos\footnote{http://mesos.apache.org/} and Vagrant\footnote{https://www.vagrantup.com/}. However, as of today, Docker is the most popular, and one of the most acknowledged containerization platforms. There is no doubt that Docker started the container revolution, and its popularity has had a steady growth. To concretize, according to previous research done in 2018, Docker was running on 20\% of hosts, and 25\% of companies had adopted Docker at that time\footnote{Numbers based on data from Datadog, a monitoring service for cloud-scale applications with thousands of companies from a big span of industries as customers.} \cite{datadog}. 
} 

Docker provides a popular registry service for the sharing of Docker images, called Docker Hub.\footnote{Docker Hub webpage: \href{https://hub.docker.com/}{\textit{https://hub.docker.com/}}} It currently hosts over 3.5 million container images, and the number keeps growing. \shorten{There are four different image repository types in Docker Hub: verified, certified, official, and community.} 
Images could be uploaded and maintained by anyone, which creates an innovative environment for anyone to contribute and participate. However, on the downside, this makes it hard for Docker to ensure that packages and applications are up to date to avoid outdated and vulnerable software.

When looking at the security of Docker, two aspects need to be considered: the security of the Docker software at the host, and the security of the Docker containers. Docker Inc. claims that ``\textit{Docker containers are, by default, quite secure; especially if you run your processes as non-privileged users inside the container.}'' \cite{docker_statement}. However, it is a simple fact that Docker (the Docker daemon and container processes) runs with root privileges by default, which exposes a huge attack surface \cite{docker_security}. A single vulnerable container is enough for an adversary to achieve privilege escalation. Hence, the security of the whole Docker ecosystem is highly related to the vulnerability landscape in Docker images.

\shorten{In the past five years, some research addressed the vulnerability landscape of Docker Hub. In 2015, BanyanOps presented results that revealed that 40\% of Docker images contained high rated vulnerabilities \cite{banyan}. Further, in 2017, Shu et al. found that on average, an image contains more than 180 vulnerabilities \cite{diva}. Then, in 2019, Socchi and Luu study revealed that the majority of official, community, and certified images hold less than 75 vulnerabilities and that the majority of verified images contain less than 180 vulnerabilities \cite{uio}. With all these findings in mind, it is clear that Docker Hub consists of a vast number of vulnerable images that could imperil a severe threat to any computer system. As the Docker Hub ecosystem is continuously changing and growing, so is doing the vulnerability landscape. This makes it important to address what the vulnerability landscape of Docker Hub is like today. 

In addition, we aim to use a different scanner, which could give a new and nuanced view on what the vulnerability landscape is like. With that said, this thesis proposes to constitute as an important contribution in relation to securing the Docker ecosystem. This is in terms of raising awareness by investigating the vulnerability landscape of Docker images.}
\vspace{-0.25cm}
\subsubsection*{Related work} \label{related_work}

One of the first to explore the vulnerability landscape of Docker Hub was BanyanOps \cite{banyan}. In 2015, they published a technical report revealing that 36\% of official images on Docker Hub contained high priority vulnerabilities \cite{banyan}. Further, they discovered that this number increases to 40\% when community images (or general images as they call it in the report) are analyzed. BanyanOps built their own vulnerability scanner based on Common Vulnerabilities and Exposures (CVE)-scores, and analyzed all official images ($\approx$75 repositories with $\approx$960 unique images) and some randomly chosen community images. However, at that time, Docker Hub consisted of just $\approx$95,000 images. 

In 2017, Shu et al. conducted a new vulnerability analysis of Docker Hub images \cite{diva}. With the aim of revealing the Docker Hub vulnerability landscape, they created their own analysis framework called DIVA (Docker image vulnerability analysis). The DIVA framework discovers, downloads, and analyses official and community images. It is based on the Clair scanner and uses random search strings to discover images on Docker Hub. 
The analysis revealed that, on average, an image (official and community) contains more than 180 vulnerabilities. They also found that many images had not been updated for hundreds of days, which is problematic from a security point of view. Further, it was observed that vulnerabilities propagate from parent to child images.

To our knowledge, the most recent vulnerability analysis of Docker Hub images was performed during spring 2019 by Socchi and Luu \cite{uio}. They investigated whether the security measures introduced by Docker Inc. (more precisely, the introduction of verified and certified image types) improved the security of Docker Hub. In addition, they inspected the distribution of vulnerabilities across repository types and whether vulnerabilities still are inherited from parent to child image. They implemented their own analyzing software using the Clair scanner, and used the results from Shu et al. \cite{diva} from 2017 as a comparison. The data set they successfully analyzed consisted of 757 images in total. Of these, 128 were official, 500 were community, 98 were verified, and 31 were certified. They only analyzed the most recent images in each repository and skipped all Microsoft repositories. Their conclusion was that the security measures introduced by Docker Inc. do not improve the overall Docker Hub security. They stated that the number of inherited vulnerabilities had dropped since the analysis of Shu et al. However, they also found that the average number of new vulnerabilities in child images had increased significantly. Further, they found that the majority of official, community, and certified repositories contain up to 75 vulnerabilities and that the majority of verified images contain up to 180 vulnerabilities.

\vspace{-0.5cm}
\subsubsection*{Our contribution}
This is an extended summary of our longer and much more detailed work \cite{WistHelsem2020}. We scrutinized the vulnerability landscape in Docker Hub images at the beginning of 2020 within the following framework:\\
    $\bullet$ Images on Docker Hub belong in one of the following four types: \textquotedbl official\textquotedbl, \textquotedbl verified\textquotedbl, \textquotedbl certified\textquotedbl, or \textquotedbl community\textquotedbl;\\
    $\bullet$ We used a quantitative mapping of the Common Vulnerability Scoring System (CVSS) \cite{cvss} (which is a numerical score indicating the severity of the vulnerability in a scale from 0.0 to 10.0) into five qualitative severity rating levels: \textquotedbl critical\textquotedbl, \textquotedbl high\textquotedbl, \textquotedbl medium\textquotedbl, \textquotedbl low\textquotedbl, or \textquotedbl none\textquotedbl \ plus one additional level \textquotedbl unknown\textquotedbl.

For performing the analysis of a significant number of images, we used an open-source vulnerability scanner tool and developed our own scripts and tools. All our developed scripts and tools are available from \cite{WistHelsem2020} and from the GitHub repository \footnote{\url{https://github.com/katrinewi/Docker-image-analyzing-tools}}.  

Our findings can be summarized as follows: \textbf{1.} The median value (when omitting the negligible and unknown vulnerabilities) is 26 vulnerabilities per image. \textbf{2.} Most of the vulnerabilities were found in the medium severity category. \textbf{3.} Around 17.8\% (430 images) do not contain any vulnerabilities, and if we are considering negligible and unknown vulnerabilities as no vulnerability, the number increase to as many as 21.6\% (523 images). \textbf{4.} As intuitively expected, when considering the average, community images are the most exposed. We found that 8 out of the top 10 most vulnerable images are community images. \textbf{5.} However, to our surprise, the certified images are the most vulnerable when considering the median value. They had the most high rated vulnerabilities as well as the most vulnerabilities rated as low. As many as 82\% of certified images contain at least either one high or critical vulnerability. \textbf{6.} Official images come out as the most secure image type. Around 45.9\% of them contain at least one critical or high rated vulnerability. \textbf{7.} The median value of the number of critical vulnerabilities in images is almost identical for all four image types. \textbf{8.} Verified and official images are the most updated, and community and certified images are the least updated. Approximately 30\% of images have not been updated for the last 400 days. \textbf{9.} There is no correlation between the number of vulnerabilities and the evaluated image features (i.e., the number of pulls, the number of stars, and the last update time). However, the images with many vulnerabilities generally have few pulls and stars. \textbf{10.} Vulnerabilities in the Lodash library and vulnerabilities in Python packages are the most frequent and most severe. The top five most severe vulnerabilities are coming from two of the most popular scripting languages, JavaScript and Python. \textbf{11.} Vulnerabilities related to execution of code and overflow are the most frequently found critical vulnerabilities. \textbf{12.} The most vulnerable package is the \texttt{jackson-databind-2.4.0} package, with overwhelming 710 critical vulnerabilities, followed by \texttt{Python-2.7.5} with 520 critical vulnerabilities.

Last but not least, when put in comparison with the three previous similar studies \cite{banyan,diva,uio}, our results are summarized in Table \ref{tab:compare}. Note that some of the cells are empty due to differences in methodologies and types of images when the studies were performed. 
\begin{table}
\centering
  \begin{tabular}{|l||@{}r@{}|@{}r@{}||@{}r@{}|@{}r@{}||@{}r@{}|@{}r@{}||@{}r@{}|@{}r@{}||}
    \noalign{\hrule height 1pt}
    \multirow{2}{*}{\parbox{1.4cm}{\centering Image type}} &
      \multicolumn{2}{@{}c@{}||}{\ 2015 \cite{banyan} \ } &
      \multicolumn{2}{@{}c@{}||}{\ 2017 \cite{diva} \ } &
      \multicolumn{2}{@{}c@{}||}{\ 2019 \cite{uio} \ } &
      \multicolumn{2}{@{}c@{}||}{\textbf{2020}} \\
    & \parbox{0.7cm}{\ vuln\vspace{0.8mm}} & \parbox{0.7cm}{\ avg} & \parbox{0.7cm}{\ vuln\vspace{0.8mm}} & \parbox{0.7cm}{\ avg} & \parbox{0.7cm}{\ vuln\vspace{0.8mm}} & \parbox{0.7cm}{\ avg} & \parbox{0.7cm}{\textbf{\ vuln}\vspace{0.8mm}} & \parbox{0.7cm}{\textbf{\ avg}}\\
    \noalign{\hrule height 1pt}
    Official & \parbox{0.6cm}{\ 36\%} & -\ \  &\ 80\% &\ 75\  & -\ \  & 170\ & \textbf{46\%} & \textbf{70} \\
    \hline
    Community & \parbox{0.6cm}{\ 40\%} & -\ \  &\ 80\% &\ 180\  & -\ \  & 150\ & \textbf{68\%} & \textbf{150} \\
    \hline
    Verified & -\ \  & - & - & - & -\ \  & 150\ & \textbf{57\%} & \textbf{90} \\
    \hline
    Certified & -\ \  & - & - & - & -\ \ & 30\ & \textbf{82\%} & \textbf{90}\\
    \noalign{\hrule height 1pt}
  \end{tabular}
  \caption{\label{tab:compare}A summary comparison table of results reported in 2015\cite{banyan}, in 2017 \cite{diva} in 2019\cite{uio} and in our work (2020). The sub-columns \textquotedbl vuln\textquotedbl \ contain the percentage of images with at least one high rated vulnerability and the \textquotedbl avg\textquotedbl \ sub-columns contain the average number of vulnerabilities found in each image type.}
\end{table}

\section{Preliminaries} 
\vspace{-0.3cm}
Virtualization is the technique of creating a virtual abstraction of some resources to make multiple instances run isolated from each other on the same hardware \cite{virtualization_how}. There are different approaches to achieve virtualization. One approach is using Virtual Machines (VMs). A VM is a virtualization of the hardware at the host. Hence, each VM has its own kernel, and in order to manage the different VMs, a software called hypervisor is required. The hypervisor emulates the Central Processing Unit (CPU), storage, and Random-Access Memory (RAM), among others, for each virtual machine. This allows multiple virtual machines to run as separate machines on a single physical machine.

\shorten{
\begin{figure*}[ht]
\centering
\begin{subfigure}{.5\textwidth}
  \centering
  \includegraphics[width=.6\linewidth]{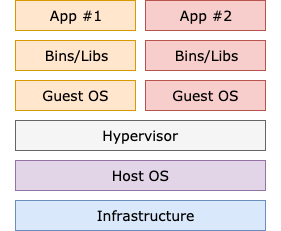}
  \caption{Virtual machine}
  \label{fig:virtual_machine}
\end{subfigure}%
\begin{subfigure}{.5\textwidth}
  \centering
  \includegraphics[width=.6\linewidth]{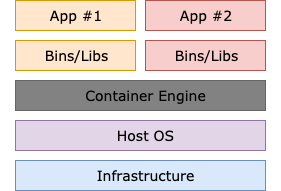}
  \caption{Container}
  \label{fig:container}
\end{subfigure}
\caption{Structure of virtual machine and container}
\label{fig:structure}
\end{figure*}
} 

In contrast to VMs, containers virtualize the Operating System (OS) level. Every container running on the same machine shares the same underlying kernel, where only bins, libraries, and other run time components are executed exclusively for a single container. In short, a container is a standardized unit of software that contains all code and dependencies \cite{container}. Thus, containers require less memory and achieve a higher level of portability than VMs. Container technology has simplified the software development process as the code is portable, and hence what is run in the development department will be the same as what is run in the production department \cite{about_docker}. 


On the Docker Hub, image repositories are divided into different categories. Repositories are either private or public and could further be either \textit{official}, \textit{community} or a \textit{verified} repository. In addition, repositories could be certified, which is a subsection of the verified category. The official repositories are maintained and vetted by Docker. Docker vets the verified ones that are developed by third-party developers. Besides being verified, certified images are also fulfilling some other requirements related to quality, support, and best practices \cite{image_types}. Community images could be uploaded and maintained by anyone. The distribution of the image repository types on Docker Hub can be seen in Table \ref{tab:dockerhub_images}. The community repository category is by far the most dominant one and makes up to $\approx$99\% of all Docker Hub repositories.

\begin{table}[]
\normalsize
\centering
\begin{tabular}{l|r}
\noalign{\hrule height 1pt}
\textbf{Repository type} & \textbf{Quantity} \\ \hline
\noalign{\hrule height 1pt}
Official        & 160                 \\ \hline
Verified        & 250                 \\ \hline
Certified       & 51                  \\ \hline
Community       & 3,064,454           \\ \hline
\noalign{\hrule height 1pt}
\textbf{Total}  & 3,064,915  \\ \hline
\noalign{\hrule height 1pt}
\end{tabular}
\caption{\label{tab:dockerhub_images}Repository type distribution on Docker Hub (February 3rd, 2020}
\end{table}

\subsection{Vulnerability databases and categorization method}
 
\vspace{-0.3cm}
The severity of vulnerabilities depends on a variety of different variables, and it is highly complex to compare them due to the diversity of different technologies and solutions. Already in 1997, the National Vulnerability Database (NVD) started working on a database that would contain publicly known software vulnerabilities to provide a means of understanding future trends and current patterns \cite{nvd_book}. The database can be useful in the field of security management when deciding what software is safe to use and for predicting whether or not software contains vulnerabilities that have not yet been discovered.

\vspace{-0.3cm}
\subsubsection*{Common Vulnerabilities and Exposures (CVE)}\label{sec:cve}

National Vulnerability Database (NVD) contains Common Vulnerabilities and Exposures (CVE) entries and provides details about each vulnerability like vulnerability overview, Common Vulnerability Scoring System (CVSS), references, Common Platform Enumeration (CPE) and Common Weakness Enumeration (CWE) \cite{cve_classification}.

CVE is widely used as a method for referencing security vulnerabilities that are publicly known in released software packages. At the time of writing, there were 130,094 entries in the CVE list.\footnote{The number of entries in the CVE list was retrieved 28. Jan 2020 from the official website:  \textit{https://cve.mitre.org}} The CVE list was created by MITRE Corporation\footnote{MITRE Corporation is a non-profit US organization with the vision to resolve problems for a safer world: \textit{https://www.mitre.org} } in 1999, whose role is to manage and maintain the list. They work as a neutral and unbiased part in order to serve in the interest of the public. Examples of vulnerabilities found in CVE are frequent errors, faults, flaws, and loopholes that can be exploited by a malicious user in order to get unauthorized access to a system or server. The loopholes can also be used as propagation channels for viruses and worms that contain malicious software \cite{cve_reference}. Over the years, CVE has become a recognized building block for various vulnerability analysis and security information exchange systems, much because it is continuously maintained and updated, and because the information is stored with accurate enumeration and orderly naming.


\begin{figure}[h!]
    \centering
    \includegraphics[width=0.6\linewidth]{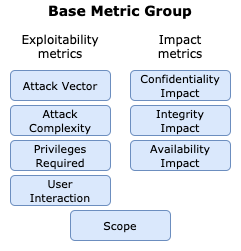}
    \caption{ Common Vulnerability Scoring System structure \cite{cvss}}
    \label{fig:cvss}
    \vspace{-0.3cm}
\end{figure}

\subsubsection*{Common Vulnerability Scoring System (CVSS)} \label{cvss}
The Common Vulnerability Scoring System (CVSS) score is a numerical score indicating the severity of the vulnerability on a scale from zero to 10, based on a variety of metrics. The metrics are divided into three metric groups: Base Metric Group, Temporal Metric Group, and Environmental Metric group. A \textit{Base Score} is calculated by the metrics in the Base Metric Group, and is independent of the user environment and does not change over time. The Temporal Metrics take in the base score and adjusts it according to factors that do change over time, such as the availability of exploit code \cite{cvss}. Environmental Metrics adjust the score yet again, based on the type of computing environment. This allows organizations to adjust the score related to their IT assets, taking into account existing mitigations and security measures that are already in place in the organization. 

In our analysis, it would not make sense to take into account the Temporal or Environmental Metrics as we wanted to discuss the vulnerability landscape independently of the exact time and environment. Therefore, only the Base Metric group will be described in more detail. It is composed of two sets of metrics: the Exploitability metrics and the Impact metrics, as can be seen in Figure \ref{fig:cvss} \cite{cvss}. The first set takes into account \textit{how} the vulnerable component can be exploited and includes attack vector and complexity, what privileges are required to perform the attack, and whether or no user interaction is required. The latter set reflects on the \textit{consequence} of a successful exploit and what impact it has on the confidentiality, integrity, and availability of the system. The last metric is \textit{scope}, which considers if the vulnerability can propagate outside the current security scope.

When the Base Score of a vulnerability is calculated, the eight different metrics from Figure \ref{fig:cvss} are being considered. Each metric is assigned one out of two to four different values, which is used to generate a vector string. The vector string is then used to calculate the Common Vulnerability Scoring System (CVSS) score, which is a numerical value between 0 and 10. In many cases, it is more beneficial to have a textual value than a numerical value. The CVSS score can be mapped to qualitative ratings where the severity is categorized as either critical, high, medium, low, or none, as can be seen in Table \ref{tab:cvss} \cite{cvss}.

\begin{table}[]
\normalsize
\centering
\begin{tabular}{l|r}
\textbf{Rating} & \textbf{CVSS Score} \\ \hline
\noalign{\hrule height 1pt}
None            & 0.0                 \\ \hline
Low             & 0.1 - 3.9           \\ \hline
Medium          & 4.0 - 6.9           \\ \hline
High            & 7.0 - 8.9           \\ \hline
Critical        & 9.0 - 10.0          \\ \hline
\noalign{\hrule height 1pt}
\end{tabular}
\caption{\label{tab:cvss}CVSS Severity scores}
\vspace{-0.3cm}
\end{table}

\section{Docker Hub vulnerability landscape}\label{rq1} 
\shorten{
This section will present the results gathered to answer \textit{\textbf{RQ1:} How can vulnerabilities found in Docker images be systemized in order to investigate the current vulnerability landscape of Docker Hub?} 
} 

\subsection{The distribution of vulnerabilities in each severity category}\label{dist_sev_vuln}
\vspace{-0.3cm}
To determine what the current vulnerability landscape is like in Docker Hub, the number of vulnerabilities found in each severity category is presented in figure \ref{fig:distribution_vulns}. As it is interesting to see how many vulnerabilities that are found in total (figure \ref{fig:severity_levels_dist}) and how many unique vulnerabilities (figure \ref{fig:unique_severity_levels_dist}) there are, both these results are presented in this section.

In figure \ref{fig:severity_levels_dist}, the results are based on vulnerability scanning of the complete data set, meaning that this result is based on all found vulnerabilities. The same vulnerability could potentially have multiple entries in the result. This is because a particular vulnerability could be found in multiple images and a single image could contain the same vulnerability in multiple packages. In figure \ref{fig:unique_severity_levels_dist}, only unique vulnerabilities are shown. However, some vulnerabilities are present in several severity categories, depending on which image it is found in. In cases like this, all versions of the vulnerability is included, which makes up a total of 14,031 vulnerabilities. 

\begin{figure*}
\centering
\begin{subfigure}{.5\textwidth}
  \centering
  \includegraphics[width=1\linewidth]{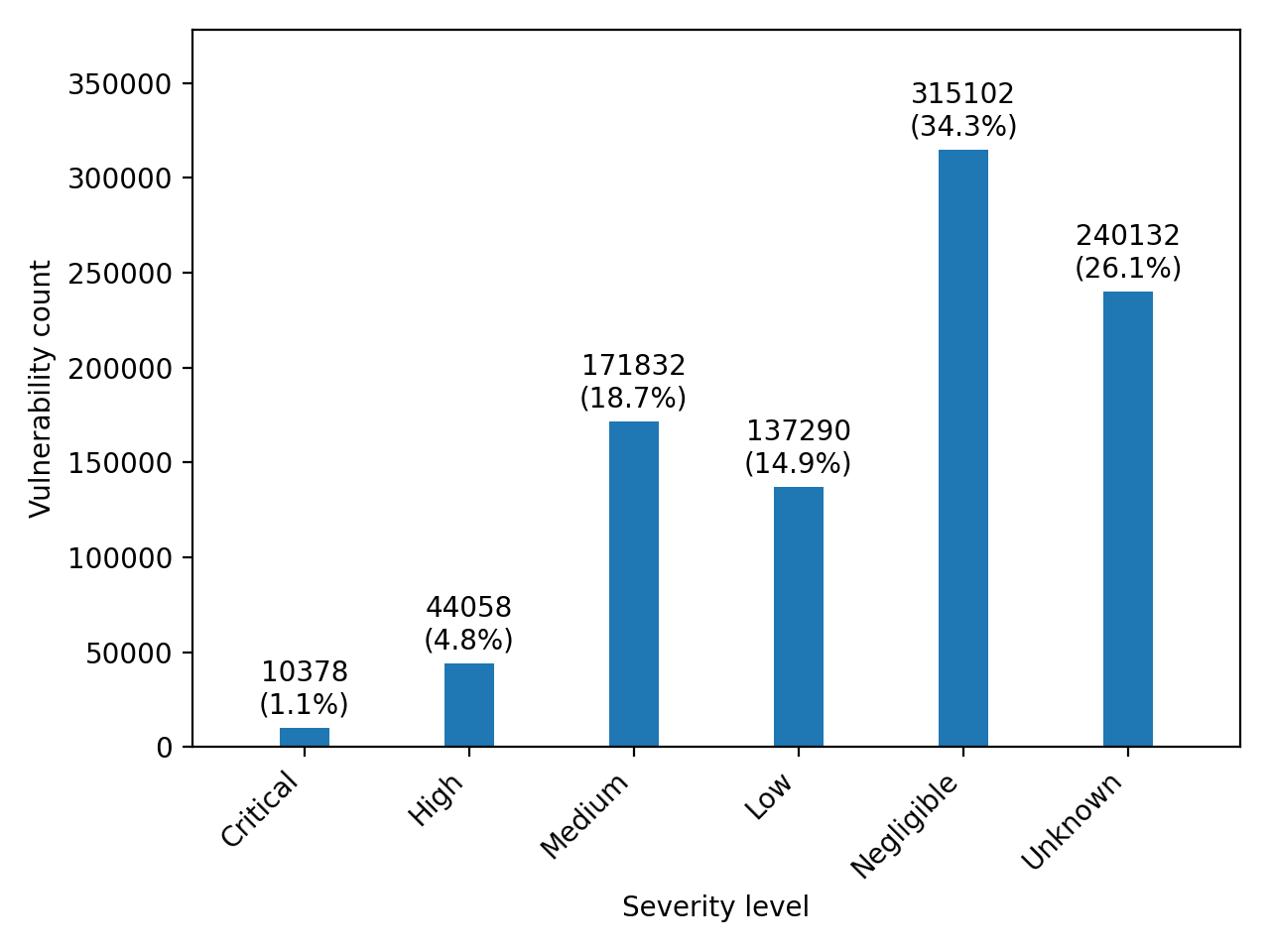}
  \caption{Distribution of all 918,792 vulnerabilities}
  \label{fig:severity_levels_dist}
\end{subfigure}%
\begin{subfigure}{.5\textwidth}
  \centering
  \includegraphics[width=1\linewidth]{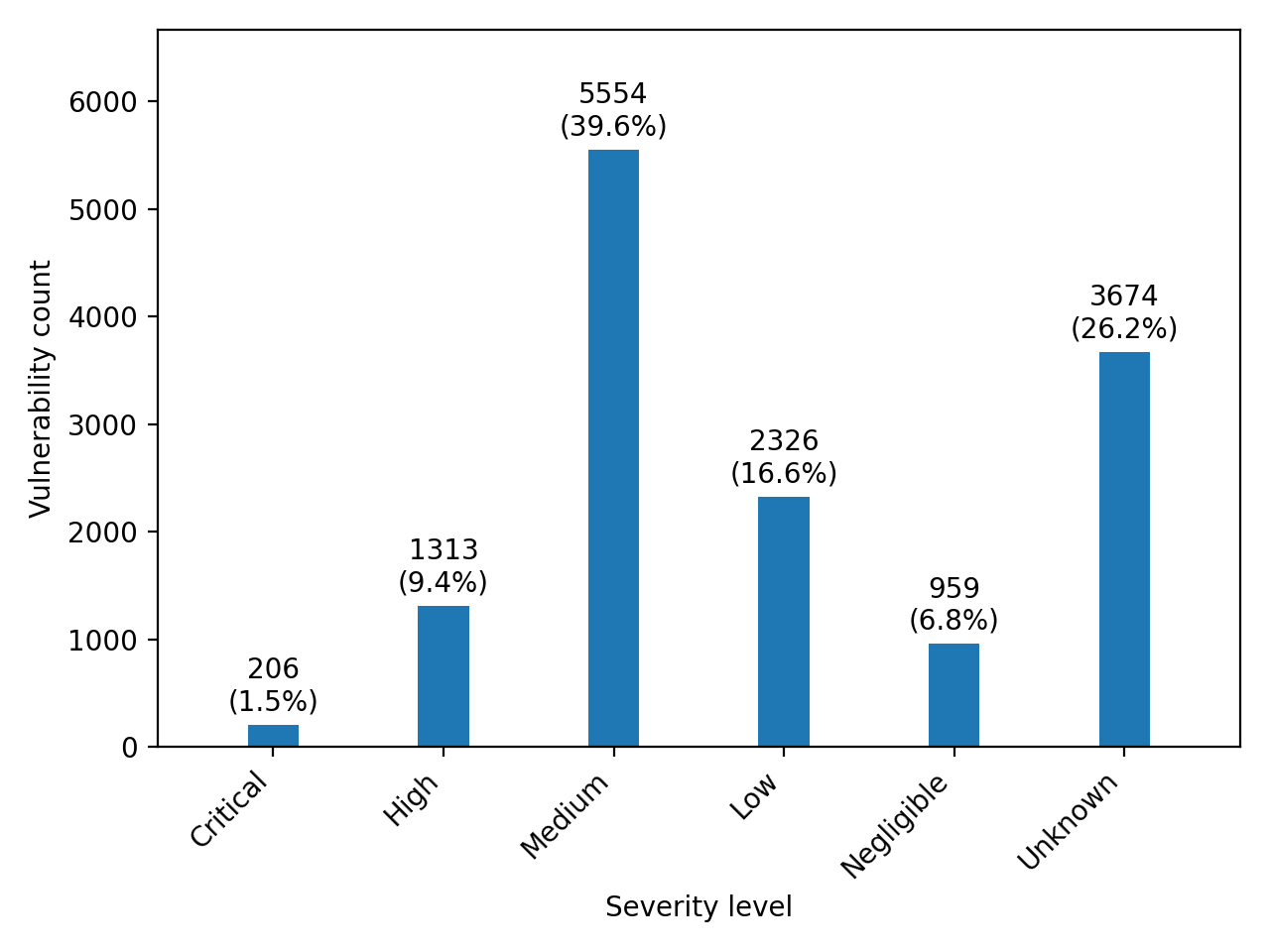}
  \caption{Distribution of 14,032 unique vulnerabilities}
  \label{fig:unique_severity_levels_dist}
\end{subfigure}
\caption{Vulnerability distribution in severity levels}
\label{fig:distribution_vulns}
\end{figure*}

\shorten{
\begin{figure*}
\centering
\begin{subfigure}{.5\textwidth}
  \centering
  \includegraphics[width=1\linewidth]{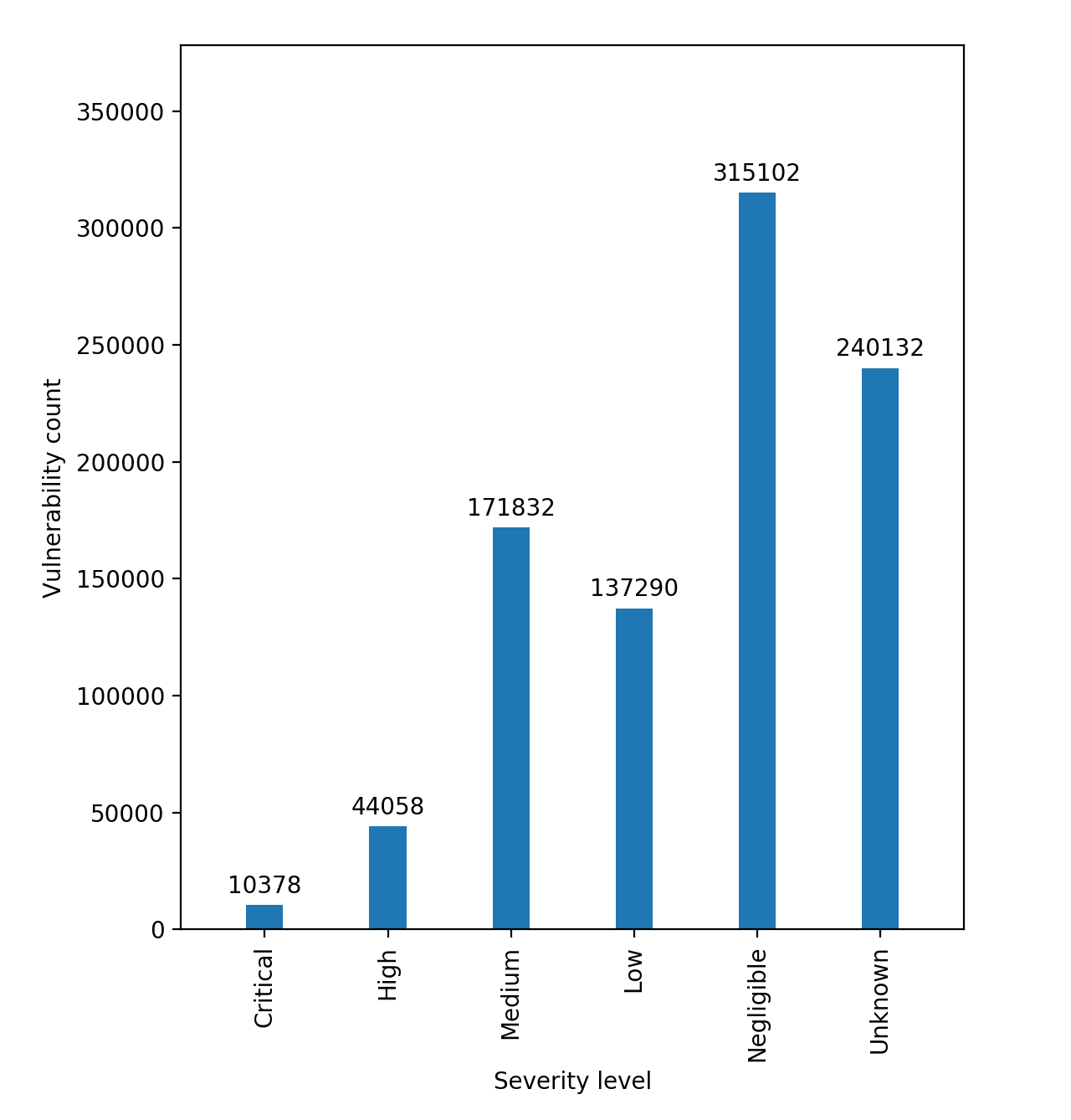}
  \caption{All vulnerabilities}
  \label{fig:severity_levels_dist}
\end{subfigure}%
\begin{subfigure}{.5\textwidth}
  \centering
  \includegraphics[width=1\linewidth]{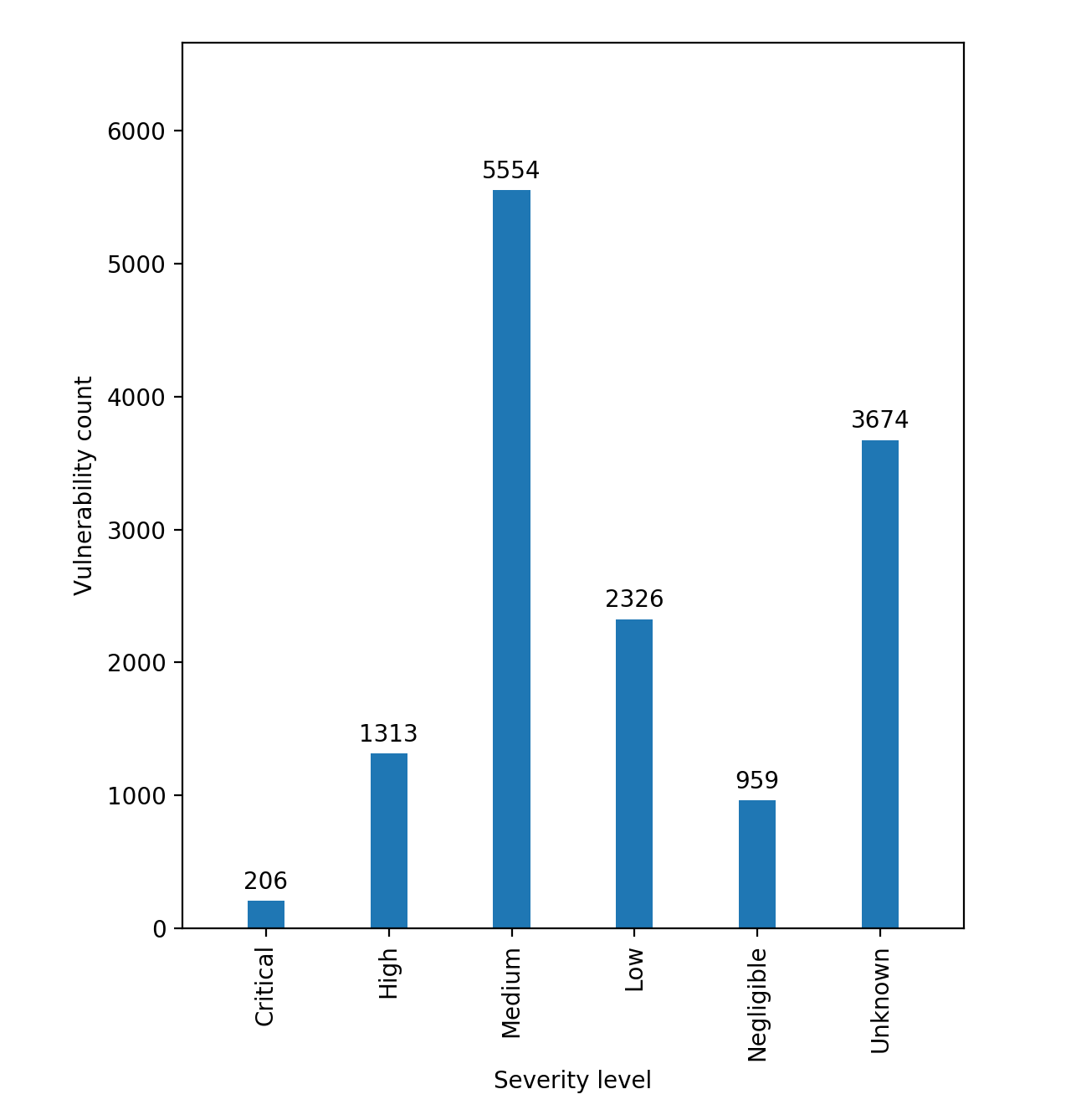}
  \caption{Unique vulnerabilities}
  \label{fig:unique_severity_levels_dist}
\end{subfigure}
\caption{Vulnerability distribution in severity levels}
\label{fig:distribution_vulns}
\end{figure*}
} 

In figure \ref{fig:severity_levels_dist}, the negligible and unknown categories clearly stands out, with a total of 315,102 and 240,132 vulnerabilities, respectively. When considering unique vulnerabilities (figure \ref{fig:unique_severity_levels_dist}), the medium category is the most dominant one with 5,554 unique vulnerabilities. When examining the relation between figure \ref{fig:severity_levels_dist} and \ref{fig:unique_severity_levels_dist}, one can observe the ratio of vulnerabilities between severity categories. It becomes clear that the negligible category contains a few number of unique vulnerabilities represented in many Docker images. Whereas the medium category has many unique vulnerabilities represented at a lower ratio. The vulnerability ratio will be explained in detail in the next paragraph.

\begin{table}[h]
\centering
\begin{tabular}{|l|r|r|r|}
\noalign{\hrule height 1pt}
\parbox{1.2cm}{\centering \vspace{1.0mm}Severity\vspace{1.0mm}}  & \parbox{2.2cm}{\centering \vspace{1.0mm}Number of vulnerabilities (A) \vspace{1.0mm}}  & \parbox{2.3cm}{\centering \vspace{1.0mm}Number of unique vulnerabilities (B) \vspace{1.0mm}}  &  \parbox{0.8cm}{\centering \vspace{1.0mm} Ratio (A/B) \vspace{1.0mm}} \\
\noalign{\hrule height 1pt} 
Critical                &10,378     &206    &50     \\ 
\hline
High                    &44,058     &1,313  &34     \\ 
\hline
Medium                  &171,832    &5,554  &31     \\ 
\hline
Low                     &137,290    &2,326  &59     \\ 
\hline
Negligible              &315,102    &959    &329    \\ 
\hline
Unknown                 &240,132    &3,674  &65     \\ 
\noalign{\hrule height 1pt}
\textbf{Total}          &918,792    &14,031 &66     \\  
\noalign{\hrule height 1pt}
\end{tabular}
\caption{\label{tab:vulns_unique_vulns}Vulnerability frequency in severity levels}
\end{table}

Table \ref{tab:vulns_unique_vulns} shows the total number of vulnerabilities, the number of unique vulnerabilities, and the ratio, measured as the total number of vulnerabilities divided by the number of unique vulnerabilities. So, for each unique vulnerability, there are a certain number of occurrences of the specific vulnerability in the data set. For example, for each unique vulnerability in the critical category, there are 50 occurrences of this vulnerability in the data set on average. For each unique negligible vulnerability, there are as many as 329 occurrences on average. This is significantly larger than the other values. Despite medium having the highest number of unique vulnerabilities, it has the lowest ratio.

\subsection{Central tendency of the vulnerability distribution}\label{central_tendency}

\vspace{-0.3cm}
We have looked at the average and median values of the number of vulnerabilities in images when disregarding the vulnerabilities that are categorized as negligible and unknown. Looking at Table \ref{tab:vulns_unique_vulns} from the previous section, one can see that negligible and unknown vulnerabilities together make up 555,234  out of the 918,792 vulnerabilities (around 60\%). As vulnerabilities in these two categories are considered to contribute with little threat when investigating the current vulnerability landscape, it gives a more accurate result to exclude these. Therefore, we calculated the average and median number of vulnerabilities in images when disregarding negligible and unknown vulnerabilities (counting them as zero). The result was 151 for the average and 26 for the median.

To investigate the data when disregarding the negligible and unknown vulnerabilities further, we created Table \ref{tab:vuln_stats} that shows statistical values of number of vulnerabilities for each image type. The results show that community images have the highest average and maximum values (158, and 6,509, respectively). The maximum value for community images is significantly larger than the average and the median, which is the case for the other three image types as well. The image type that is considered as the least vulnerable is official. It has the lowest average of 73 and the lowest median value of 9. Further, the maximum value for official images is the second lowest. The lowest maximum value belongs to certified, and is only 428. Although certified has the lowest maximum value, it has the highest median value. This indicates that a larger portion of the images have many vulnerabilities. As a final note, all four image types contain at least one image with zero vulnerabilities. 

\begin{table}[]
\centering
\begin{tabular}{|l|r|r|r|r|r|}
\noalign{\hrule height 1pt}
Image type & \parbox{1.1cm}{\centering \vspace{1.0mm}Number of analyzed images\vspace{1.0mm}} & \parbox{1.1cm}{\centering \vspace{1.0mm}Number of vulnerabilities\vspace{1.0mm}} & \parbox{0.8cm}{\centering Average} & \parbox{0.8cm}{\centering Median} & Max \\ 
\noalign{\hrule height 1pt}
Verified        &60     &6,073      &101.2  &13 &1,128   \\ \hline
Certified       &22     &1,987      &90.3   &37 &428     \\ \hline
Official        &157    &11,489     &73.2   &9  &1,615   \\ \hline
Community       &2,173  &344,009    &158.3  &28 &6,509   \\ 
\noalign{\hrule height 1pt}
\end{tabular}
\caption{\label{tab:vuln_stats}Statistical values for vulnerabilities per image type, disregarding negligible and unknown vulnerabilities.}
\end{table}

\subsection{Vulnerabilities in each image type}\label{vuln_img_type}
\begin{figure}[]
    \centering
    \includegraphics[width=0.8\linewidth]{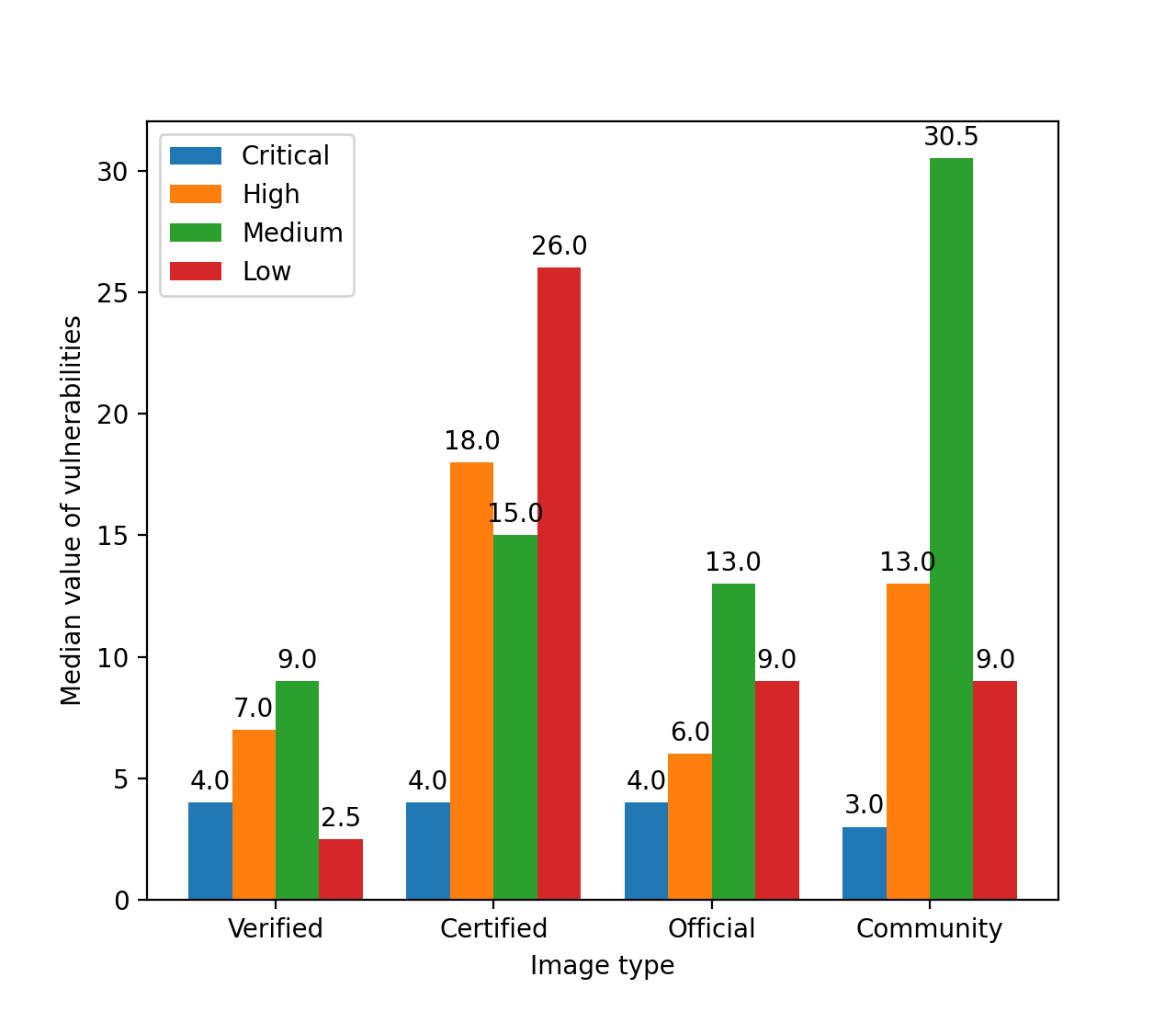}
    \caption{Median values of vulnerabilities for each severity category and image type}
    \label{fig:median_image_types}
\end{figure}

Since the median describes the central tendency better than the average when the data is skewed here we will work with the median values (given in Figure \ref{fig:median_image_types}). Note that only critical, high, medium and low vulnerabilities are included in the figure. The negligible and unknown vulnerabilities are not included here because they do not usually pose as significant threats, and therefore do not contribute with additional information when investigating the current vulnerability landscape.

The results show that the median of critical vulnerabilities is almost the same for all four image types (4.0 and 3.0). The other severity categories are more varied across the image types. The high severity category is the most represented in certified images, while the medium category is the most represented in the community images. For verified, official and community images, the medium severity has the highest median, while the certified images has the most low vulnerabilities. Overall, it is the certified images that are the most vulnerable.

\subsection{Images that contain the most critical vulnerabilities}\label{img_most_critical}
Out of all 2,412 successfully analyzed images, this section will present the most vulnerable ones. Table \ref{tab:vuln_images} displays the most vulnerable images based on the number of critical vulnerabilities in each image. In cases where the critical count is the same, the image with the highest number of high rated vulnerabilities is considered as the most vulnerable one. The \textit{Number of pulls} column denotes the total number of pulls (downloads) for each image.

\begin{table*}[]
\centering
\small\addtolength{\tabcolsep}{-5pt}
\begin{tabular}{|l|l|r|r|r|r|r|}
\noalign{\hrule height 1pt}
& \textbf{Image}  &  \textbf{Critical} & \textbf{High} & \textbf{Medium} & \textbf{Low} & \textbf{Number of pulls}\\ \noalign{\hrule height 1pt}
1 & pivotaldata/gpdb-pxf-dev   &   822  & 698  & 576 & 132 &139,246,839\\ \hline
2 & cloudera/quickstart & 571  & 2,155 & 1,897 & 158    & 6,892,856  \\ \hline
3 & silverpeas  & 341   & 264 & 397 & 226 & 828,743 \\ \hline
4 & microsoft-mmlspark-release   & 184   & 428 & 264 & 252 & 1,509,541  \\ \hline
5 & anchorfree/hadoop-slave & 168 & 636 & 797 & 107 & 5,375,424\\ \hline
6 & saturnism/spring-boot-helloworld-ui & 133  & 217 & 112 & 2 &12,686,987 \\ \hline
7 & pantsel/konga      & 133& 39 & 169 & 0 & 12,431,685\\ \hline
8 & renaultdigital/runner-bigdata-int & 127 & 335  &  691 & 103 & 4,787,745\\ \hline
9 & springcloud/spring-pipeline-m2  & 125 & 293  & 2,027 & 1,357 & 8,359,973\\ \hline
10 & raphacps/simpsons-maven-repo & 122 & 271  &  399 & 2  &36,136,733 \\ \hline
\noalign{\hrule height 1pt}
\end{tabular}
\caption{\label{tab:vuln_images}The most vulnerable images sorted by critical count}
\end{table*}

Out of the top 10 most vulnerable images, there are 8 community images, 1 official image (silverpeas) and 1 verified image (microsoft-mmlspark-release). There are big variations in the number of vulnerabilities in all presented severity levels. The most vulnerable image, \textit{pivotaldata/gpdb-pxf-dev}, has \textasciitilde250 more critical vulnerabilities than the second most vulnerable image. However, the second most vulnerable image, \textit{cloudera/quickstart}, contains as many as 2,155 high rated vulnerabilities, which is \textasciitilde1500 more vulnerabilities than the one rated as the most vulnerable image. It was chosen to focus on the critical vulnerabilities in the ranking of the most vulnerable images. This is because it is the highest possible ranking and hence the most severe vulnerabilities will be found in this category. The other severity categories are included in the table as extra information and to give a clear view on the distribution of vulnerabilities. From the number of pulls column one can observe that the most vulnerable image is also the most downloaded one out of the top 10, with as many as 139,246,839 pulls. This is approximately 100 million more pulls compared to the second most pulled image on this list (the \textit{raphacps/simpsons-maven-repo} image). There is no immediate correlation that could be observed between the number of pulls and the number of vulnerabilities in these images.

\begin{figure}[]
    \centering
    \includegraphics[width=0.7\linewidth]{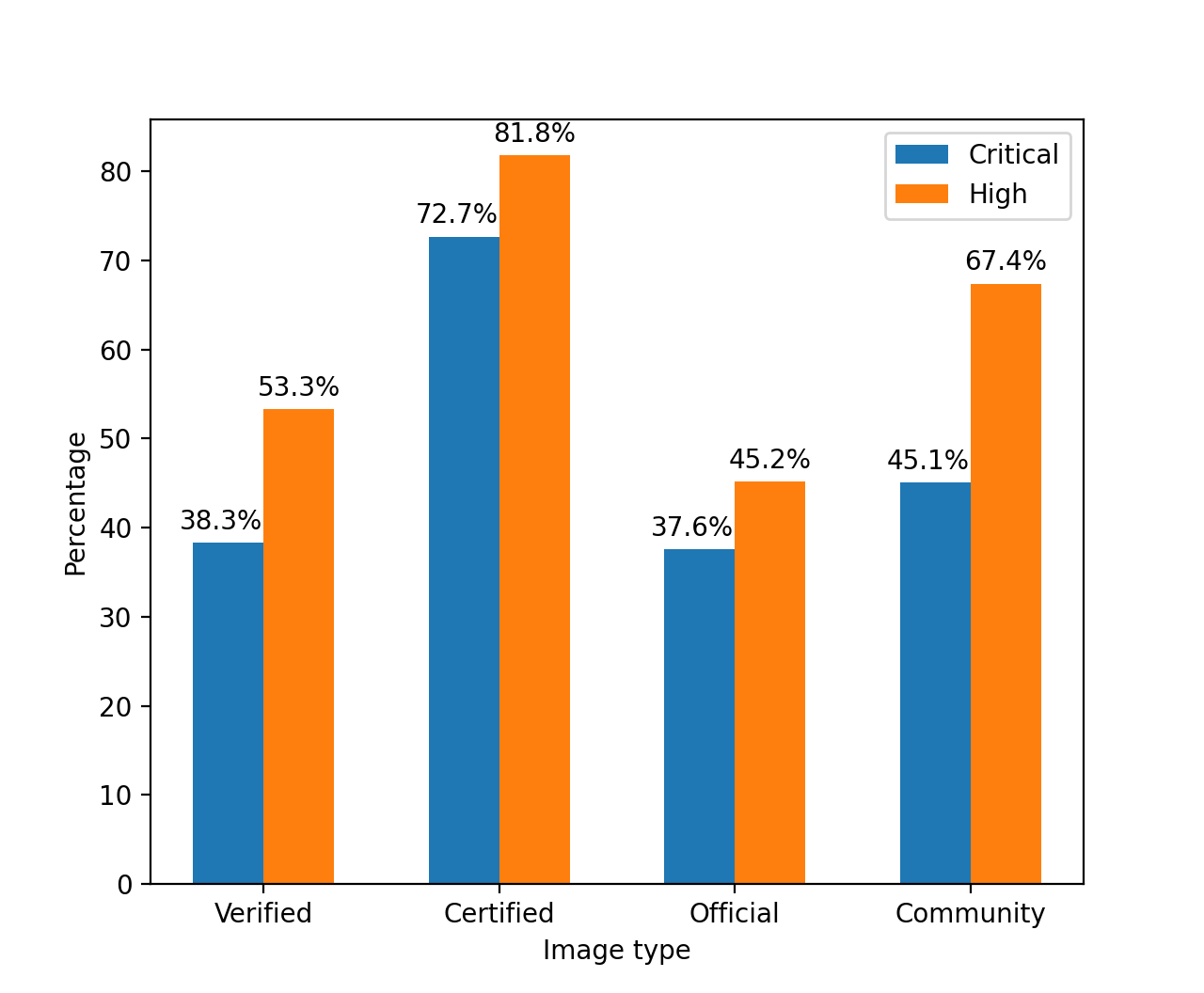}
    \caption{The percentage of images that contain at least one high or critical rated vulnerability.}
    \label{fig:critical_high_percent}
\end{figure}

\subsection{Percentage of images with critical and high vulnerabilities}\label{percentage_img}
It is enough with a single vulnerability for a system to be compromised. Thus, we determine what percentage of images that contain at least one high or critical rated vulnerability for each image type, as shown in Figure \ref{fig:critical_high_percent}.

Our results (Figure \ref{fig:critical_high_percent}) reveal that the certified image type, which is a subsection of the verified image type, is the most vulnerable by the means of this measure. 81.8\% of all certified images contain at least one vulnerability with high severity level and 72.7\% of them contain at least one critical vulnerability. Community images come out as the second most vulnerable image type. 67.4\% have high vulnerabilities and 45.1\% have critical vulnerabilities. The third most vulnerable image type is verified, followed by official.

When combining these results, to investigate what amount of the image types that contain \textit{either} at least one critical or high rated vulnerability, the results are as follows: 81.8\% for certified images, 68.4\% for community images, 56.7\% for verified images and 45.9\% for official images. This makes the official images the least vulnerable image type. However, it should be 
emphasized that still almost half of the official images contain critical or high rated vulnerabilities as presented in this section.


\begin{figure*}
\centering
\begin{subfigure}{.5\textwidth}
  \centering
  \includegraphics[width=1\linewidth]{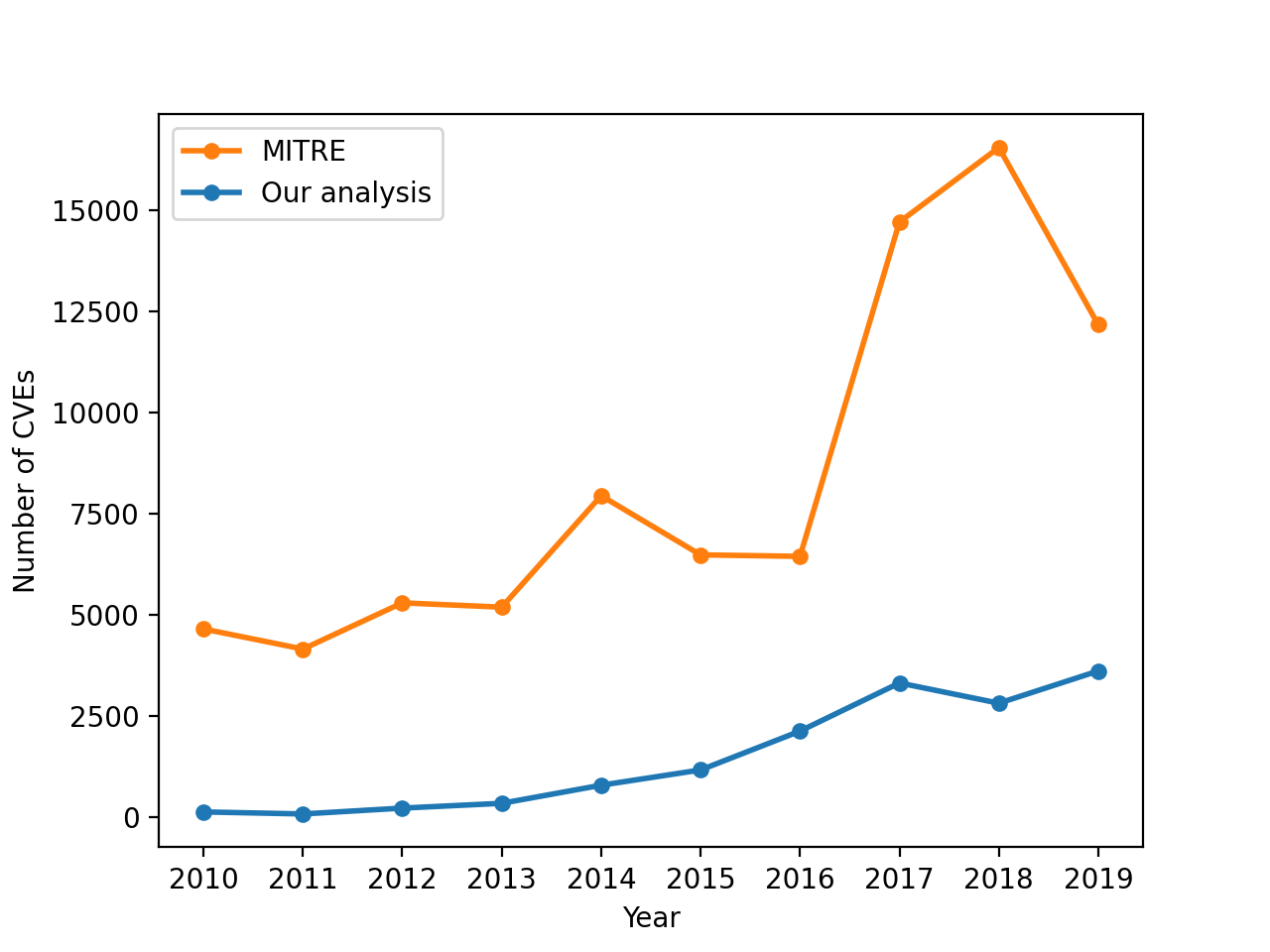}
  \caption{}
  \label{fig:cve_trend}
\end{subfigure}%
\begin{subfigure}{.5\textwidth}
  \centering
  \includegraphics[width=1\linewidth]{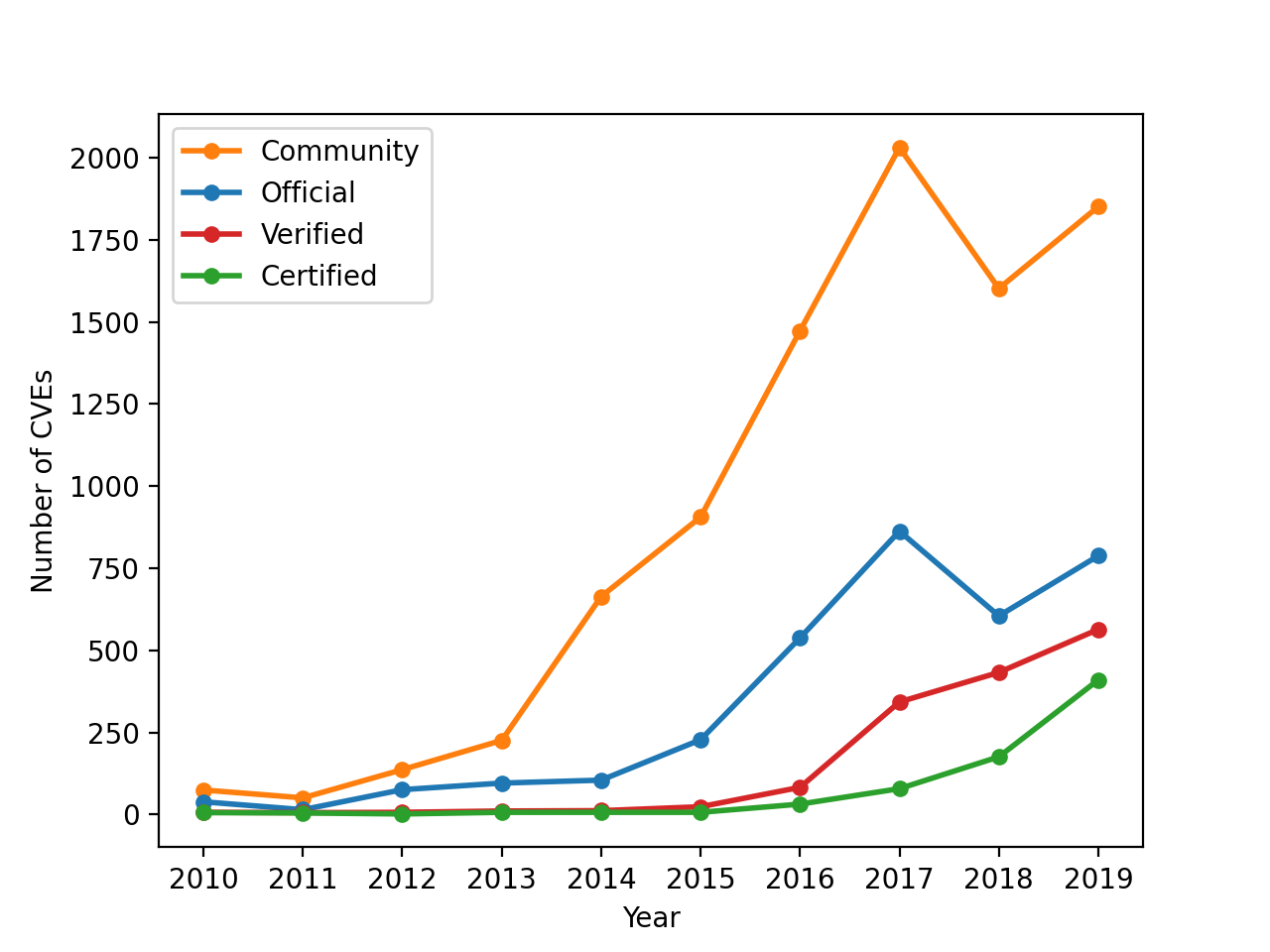}
  \caption{}
  \label{fig:cve_trend_types}
\end{subfigure}
\caption{CVE trend from 2010 to 2019, (a) displays all reported CVEs and all found, unique CVEs in our analysis, (b) displays the CVEs in the different image types from our analysis.}
\label{fig:cve_trend_fig}
\end{figure*}

\subsection{The trend in CVE vulnerabilities} \label{sec:cve_trend}
This  section  will  focus  on  the  trend  of  all  reported Common  Vulnerabilities  and  Exposures  (CVE)  vulnerabilities  each  year  compared  to  the  number  of  unique CVE  vulnerabilities  found  throughout  our  analysis.  Data gathered from the CVE Details database \cite{cve_details_data} is used to display the number of new reported Common Vulnerabilities and Exposures (CVE) vulnerabilities each year.

In Figure \ref{fig:cve_trend} the reported  Common  Vulnerabilities and  Exposures  (CVE)  vulnerabilities  each  year  is  presented together with the unique CVE vulnerabilities found in our analysis from 2010 to 2019. The orange line shows how  the  number  of  new  discovered  CVE  vulnerabilities varies by a few thousand vulnerabilities each year. However, there is a significant increase in 2017. This increase is not reflected in the data from our analysis, which is following a steady increase in the years from 2014 to 2017. This increase can be explained by the introduction of Docker Hub in 2014, making new vulnerabilities more represented in images. As a final observation, the number of new reported vulnerabilities from MITRE between 2018 and 2019 is decreasing, while there is an increase in our results. 

Figure \ref{fig:cve_trend_types} shows the number of unique vulnerabilities found in each image type (i.e. community, official, verified and certified) in our analysis from 2010 to 2019. This figure gives an insight in how the overall changes are reflected in each image type. Verified and certified images have had an increase in the number of unique Common Vulnerabilities and Exposures (CVE)  vulnerabilities each year from 2015. Community and official images, however, have had a significant decrease of unique vulnerabilities from 2017 to 2018. It is noteworthy to point out that the curves are affected by the time of introduction of the different image types. Official images were introduced in 2014, whereas verified and certified images were introduced in 2018.

\subsection{Days since last update}\label{days_since_last_update}
There is a high variation in how often Docker Hub images are updated. Intuitively, this affects the vulnerability landscape of Docker Hub. Hence, we have gathered data about when images were last updated, and calculated the number of days since the images were last updated, counting back from February 25th, 2020. The data set consists of last updated data for all analyzed images, except five.

A brief analysis of the numbers from our database revealed that 31.4\% of images have not been updated in 400 days or longer and 43.8\% have not been updated in 200 days or longer. The percentage of images that have been updated during the last 14 days are 29.8\%. This implies that if these numbers are representative for all images on Docker Hub, a third of the images (31.4\%) on Docker Hub have not been updated in the last 400 days or longer. 

To go into more detail, Table \ref{tab:days_since} presents how often images in each of the image types are updated. Community and certified images are the least updated image categories, where 47.0\% of community images and 36.4\% of certified images have not been updated for the last 200 days or more. The verified images are the most frequently updated category, where 83.3\% of images have been updated during the last 14 days.

\begin{table}[]
\small\addtolength{\tabcolsep}{-5pt}
\centering
\begin{tabular}{|l|r|r|r|}
\noalign{\hrule height 1pt}
Image type  &  \parbox{1.4cm}{\centering \vspace{1.0mm}More than 400 days\vspace{1.0mm}}  & \parbox{1.4cm}{\centering \vspace{1.0mm}More than 200 days\vspace{1.0mm}}  & \parbox{1.4cm}{\centering \vspace{1.0mm}Less than 14 days\vspace{1.0mm}} \\ 
\noalign{\hrule height 1pt}
Community        & 33.9\% & 47.0\%   & 27.0\%             \\ \hline
Official        & 9.6\% & 14.7\% & 51.3\%                \\ \hline
Certified       & 18.2\% & 36.4\% & 13.6\%                   \\ \hline
Verified       & 1.7\% & 5.0\% & 83.3\%           \\ \hline
\noalign{\hrule height 1pt}
\end{tabular}
\caption{\label{tab:days_since}The time since last update for all image types presented in percentage}
\end{table}

A handful of certified images are highly affecting the percentages from Table \ref{tab:days_since}, because the overall number of certified images is small. Official images contain a high portion of images that have been updated recently (January 2020 to March 2020), and some more spread values with images that have not been updated since 2016. The verified images are the most updated image type, where there is only one image with the last updated time earlier than May 2019.

\section{Correlation between image features and vulnerabilities} \label{sec_rq2}
We investigate whether or not the number of vulnerabilities in an image is affected by a specific image feature, such as the number of times the image has been pulled, the number of stars an image has been given, or the number of days since the image was last updated. \shorten{The following sections present the results required to answer the following question: \textit{How does image features and number of vulnerabilities correlate in images?}} 
In order to find out whether there is a correlation, we used Spearman's $r_s$ correlation coefficient \cite{lehman2005jmp}. Spearman's correlation was chosen because our data set contain skewed values and are not normally distributed. When handling entries that contained empty values, we opted for the approach of complete case analysis, which means omitting incomplete pairs. The alternative would be imputation of missing values, which means to create an estimated value based on the other data values. However, this approach was not chosen because the values of our data set are independent of each other. 

\subsubsection*{Correlation between pulls and vulnerabilities}
\begin{figure}[]
    \centering
    \includegraphics[width=1.0\linewidth]{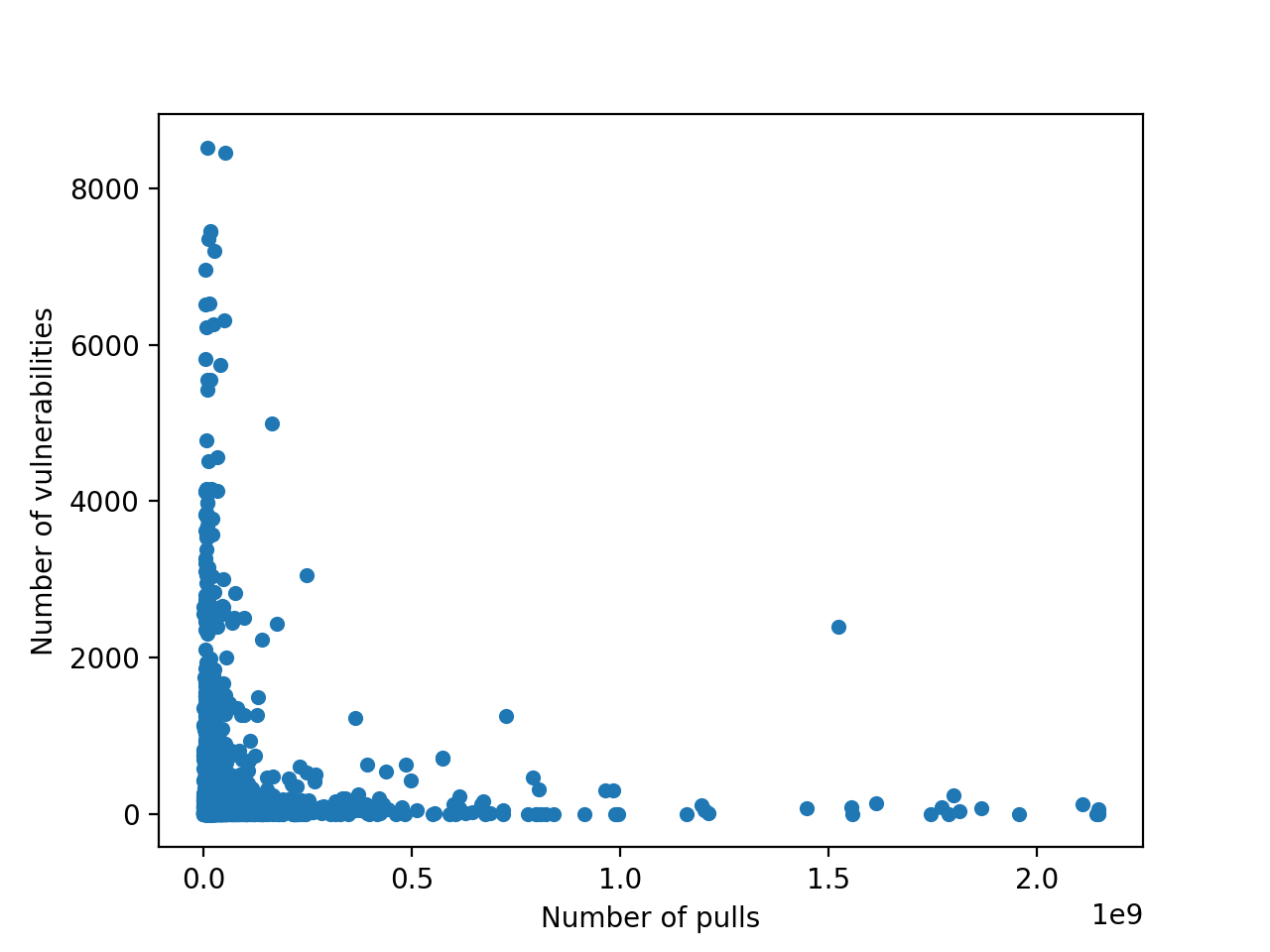}
    \caption{Number of pulls and number of vulnerabilities for each image}
    \label{fig:scatter_pulls}
\end{figure}

To check the folklore wisdom about the following correlation: \textit{images with the most pulls generally have few vulnerabilities, and images with the most vulnerabilities generally have few pulls}, we created a scatter plot given in Figure \ref{fig:scatter_pulls}. However, after calculating the Spearman correlation coefficient between the number of pulls and number of vulnerabilities for the whole set of investigated images we got $r_s = -0.1115$. This is considered as no particular correlation. To explain this, we refer to the meaning of having a high negative correlation: the markers would gather around a decreasing line (not necessarily linear), indicating that images with more pulls have less number of vulnerabilities. In the case of high positive correlation, the opposite would apply i.e. the line would be increasing. 

\shorten{While for the leftmost part of the plot (in the range $[0, 0.5\times 10^9]$ pulls) that decreasing correlation curve is visible (see the zoomed sub-plot in Figure \ref{fig:scatter_pulls}), in the range $[0.5\times 10^9, 2.5\times 10^9]$ there is no such a decreasing trend line indicating that for images with less than 1000 vulnerabilities, there are little to no correlation between the number of pulls and the number of vulnerabilities.
} 

\FloatBarrier

\subsubsection*{Correlation between stars and vulnerabilities}
\begin{figure}[h]
    \centering
    \includegraphics[width=1.0\linewidth]{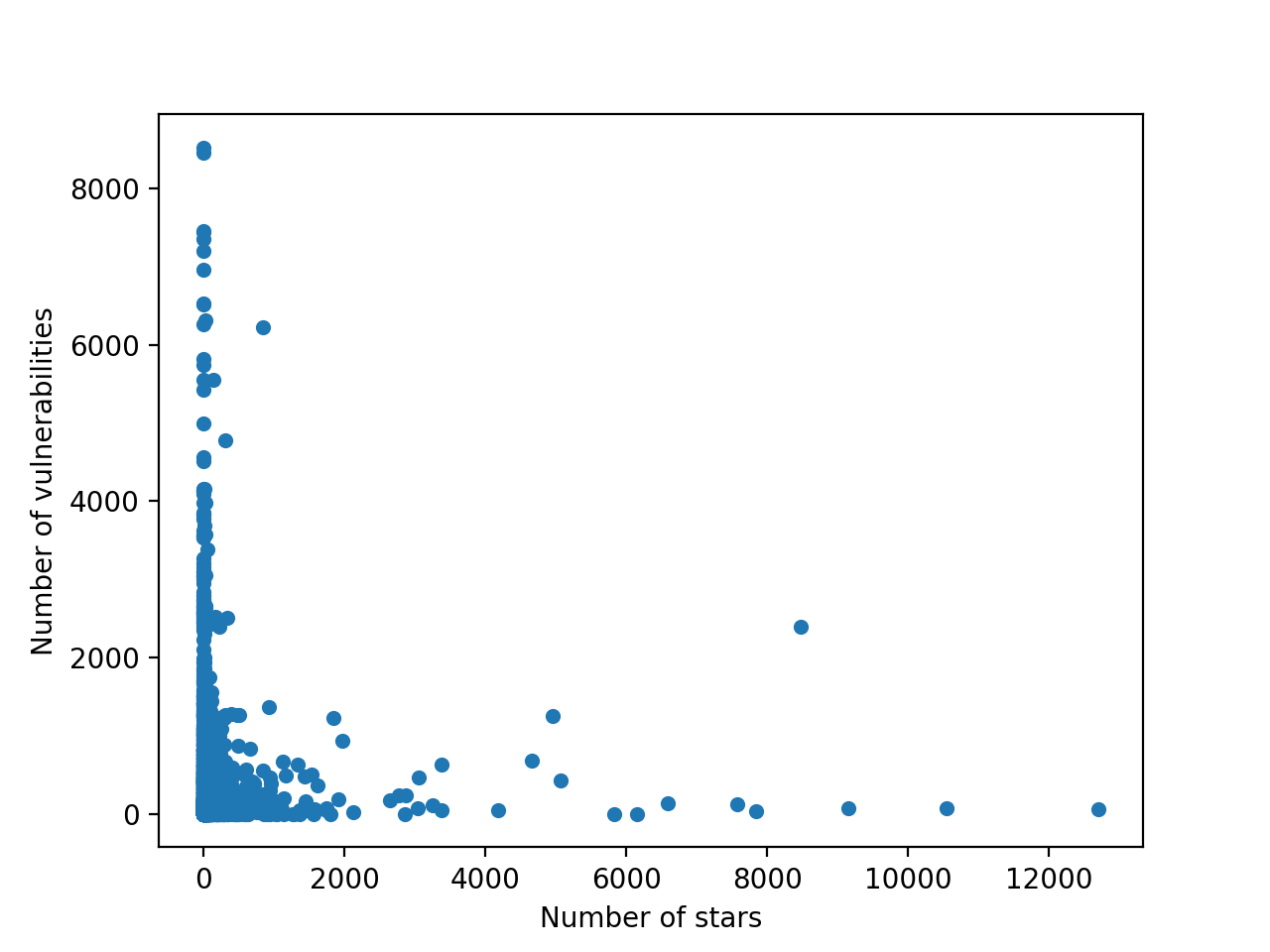}
    \caption{Number of stars and number of vulnerabilities for each image}
    \label{fig:scatter_stars}
\end{figure}

The correlation coefficient between the number of stars and number of vulnerabilities is $r_s = -0.0335$. Figure \ref{fig:scatter_stars} shows the scatter plot when including number of stars instead of number of pulls. The plot is similar to Figure \ref{fig:scatter_pulls}, but the correlation is even weaker.
\FloatBarrier

\subsubsection*{Correlation between time since last update and vulnerabilities}
\begin{figure}[]
    \centering
    \includegraphics[width=1.0\linewidth]{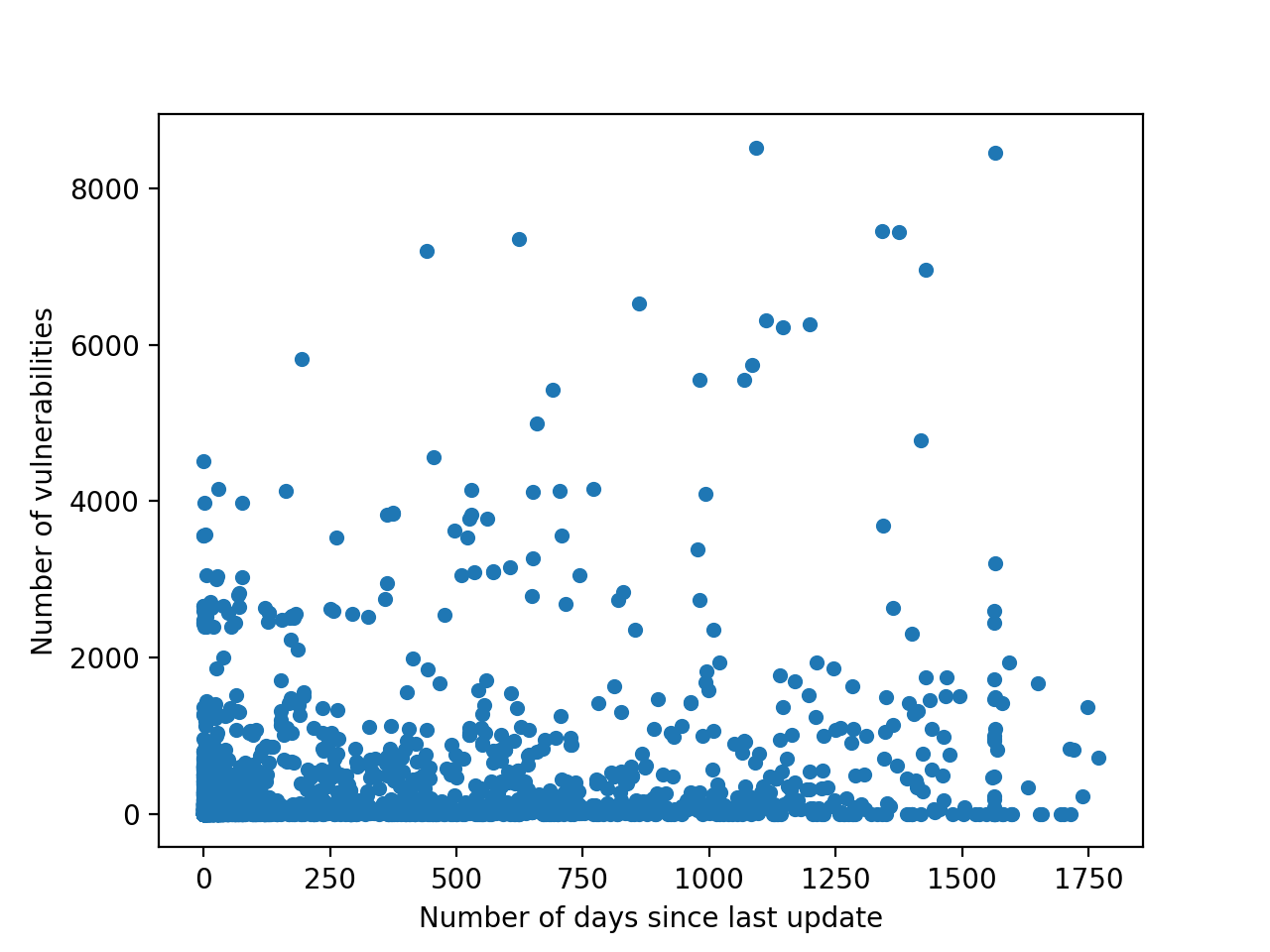}
    \caption{Number of days since last update and number of vulnerabilities for each image}
    \label{fig:scatter_days}
\end{figure}

This correlation is calculated by computing the number of days since the last update counting from the day we gathered the data (which was February 25, 2020). The correlation was $r_s = 0.1075$, which shows a positive correlation as opposed to the other two. Figure \ref{fig:scatter_days} shows the scatter plot, and although the markers are approaching an increasing line a tiny bit, this is minimal. The value of 0.1075 is still not enough to state that there is a strong correlation between the number of vulnerabilities and time since the last update. The markers slightly approach an increasing line, indicating a weak tendency that there are more vulnerabilities in images that have not been updated for a long time. Still, the distribution of markers is relatively even along the x-axis with the most markers in the lower part of the y-axis, supporting that there is no correlation.
\FloatBarrier

\section{The most severe vulnerabilities} \label{severe_vuln}
\shorten{
This section will present which vulnerabilities that are the most severe, as stated in \textbf{RQ3:} \textit{Which types of vulnerabilities are the most severe?} }

 
\subsection{The most represented critical vulnerabilities} \label{most_rep_vuln}
The most represented severe vulnerabilities are, intuitively, the ones having the highest impact on the vulnerability landscape. Table \ref{tab:vulns} presents the most represented critical rated vulnerabilities in descending order. The results are obtained by counting the number of occurrences for each vulnerability ID in the critical severity level. The critical count column is the number of occurrences for a specific vulnerability. Lastly, the type(s) column presents the vulnerability type of each of the vulnerabilities. This data is gathered from the CVE Details database \cite{cve_details}.

\begin{table}[h]
\small\addtolength{\tabcolsep}{-2pt}
\centering
\begin{tabular}{|r|l|c|l|}
\noalign{\hrule height 1pt}
\textbf{}  &  \parbox{1.3cm}{ \vspace{1.0mm}\textbf{Vulnerability ID}\vspace{1.0mm}} & \parbox{1.0cm}{\vspace{1.0mm}\textbf{Critical count}\vspace{1.0mm}}  & \textbf{Type(s)} \\ 
\noalign{\hrule height 1pt}
  1 &   CVE-2019-10744  &    466 & \parbox{3.2cm}{ \vspace{1.0mm}Improper Input \\ Validation\vspace{1.0mm}}  \\ \hline
2  & CVE-2017-1000158  &    464   & \parbox{3.2cm}{\vspace{1.0mm}Execute Code, Overflow\vspace{1.0mm}}    \\ \hline
3  & CVE-2019-9948 &   378 & \parbox{3.2cm}{\vspace{1.0mm}Bypass a restriction or \\ similar\vspace{1.0mm}} \\ \hline
4   & CVE-2019-9636   & 374  & \parbox{3.2cm}{\vspace{1.0mm}Credentials Management Errors\vspace{1.0mm}} \\ \hline
5   &  CVE-2018-16487 & 365 & \parbox{3.2cm}{\vspace{1.0mm}Security Features\vspace{1.0mm}} \\ \hline
6  & CVE-2018-14718   & 354 & \parbox{3.2cm}{\vspace{1.0mm}Execute Code\vspace{1.0mm}} \\ \hline
7    & CVE-2018-11307    & 337 & \parbox{3.2cm}{\vspace{1.0mm}Deserialization of \\ Untrusted Data\vspace{1.0mm}} \\ \hline
8 & CVE-2018-7489  &  318 & \parbox{3.2cm}{\vspace{1.0mm}Execute Code, Bypass a restriction or similar\vspace{1.0mm}}\\ \hline
9  & CVE-2016-5636  & 302 & Overflow \\ \hline
10   & CVE-2017-15095  &  295 &	Execute Code \\ \hline
\noalign{\hrule height 1pt}
\end{tabular}
\caption{\label{tab:vulns}The most represented vulnerabilities (based on critical severity level).}
\vspace{-0.5cm}
\end{table}

\subsection{Vulnerability characteristics} \label{vuln_charac}
We elaborate the top five most represented vulnerabilities presented in Table \ref{tab:vulns} regarding their characteristics and common features\footnote{Information about all vulnerabilities could be found by visiting https://nvd.nist.gov/vuln/detail/}. The top five severe vulnerabilities are coming from two most popular script languages: JavaScript and Python. As a general observation, the execute code is the most common vulnerability type, followed by overflow.

The most represented critical vulnerability is found 466 times throughout our scanning. It has vulnerability ID \textit{CVE-2019-10744}, and a base score of 9.8, which is in the upper range of the critical category (to examine how base scores are determined, see Section \ref{cvss}). The vulnerability is related to the JavaScript library lodash, which is commonly used as a utility function provider in relation to functional programming. This particular vulnerability is related to improper input validation and makes the software vulnerable to prototype pollution. It is affecting versions of lodash lower than 4.17.12 \cite{vuln1}. In short, this means that it is possible for an adversary to execute arbitrary code by modifying the properties of the Object.prototype. This is possible as most JavaScript objects inherit the properties of the built in Object.prototype object. The fifth vulnerability on the list, \textit{CVE-2018-16487}, is also related to lodash and the prototype pollution vulnerability.

Further, the second, third and fourth most represented critical vulnerabilities are related to Python vulnerabilities. The second vulnerability with vulnerability ID, \textit{CVE-2017-1000158}, is related to versions of Python up to 2.7.13. The base score is rated 9.8, and the vulnerability enables arbitrary code execution to happen through an integer overflow leading to a heap-based buffer overflow \cite{vuln2}. Overflow vulnerabilities could be of different types, for instance heap overflow, stack overflow and integer overflow. Heap overflow and stack overflow are related to overflowing a buffer, whereas integer overflow could lead to a buffer overflow. A buffer overflow is related to overwriting a certain allocated buffer, causing adjacent memory locations to be overwritten. Any exploit of these kinds of vulnerabilities are typically related to the execution of arbitrary code, where the adversary is taking advantage of the buffer overflow vulnerability to run malicious code. 

The third presented vulnerability with vulnerability ID \textit{CVE-2019-9948} is affecting the Python module urllib in Python version 2.x up to 2.7.16. It is rated with 9.1 as base score. This vulnerability makes is easier to get around security mechanisms that blacklist the \texttt{file:URIs} syntax, which in turn could give an adversary access to local files such as the \textit{/etc/passwd} file \cite{vuln3}. The fourth vulnerability is found 374 times and has vulnerability ID \textit{CVE-2019-9636}. It is affecting both the second and the third version of Python (versions 2.7.x up to 2.7.16, and 3.x up to 3.7.2). This vulnerability is also related to the urllib module, more precisely, incorrect handling of unicode encoding. The result is that information could be sent to different hosts than intended if it was parsed correctly \cite{vuln4}. It has a base score of 9.8.

\section{Vulnerabilities in packages} \label{rq4} 
\shorten{To investigate the origin of the most severe vulnerabilities, we seek to determine the top 10 most vulnerable packages, as well as how many images that use these packages. We will also determine the vulnerabilities in the most used packages. This is in accordance with \textbf{RQ4:} \textit{Which packages contain the most severe vulnerabilities?} }

\subsection{The most vulnerable packages} \label{sec:most_vulnerable_packages}
Table \ref{tab:vuln_packages} presents the packages that contain the most critical vulnerabilities. The critical count column is obtained by counting the total number of occurrences of critical vulnerabilities in each package, while the image count column is the number of images that uses each package.
\shorten{
\subsubsection{Number of images that use the most vulnerable packages}
To be able to get a better view on the security impact each of these packages have, Figure \ref{fig:packages} is presented. It displays the number of images that uses each of the packages from Table \ref{tab:vuln_packages}. 
}

There is a clear relation between the most vulnerable packages and the most represented vulnerabilities (Section \ref{severe_vuln}), as expected. For example, vulnerabilities found in Python version 2.x packages and in the Lodash package are both presented in Section \ref{severe_vuln}.
\begin{table}[]
\centering
\begin{tabular}{rlr|r}
\noalign{\hrule height 1pt}
  & \textbf{Package}  &   \textbf{Critical count} &   \textbf{Image count} \\ \noalign{\hrule height 1pt}
1 & jackson-databind-2.4.0      &   710  & 15\\ \hline
2 & Python-2.7.5        & 520    & 207   \\ \hline
3 & jackson-databind-2.9.4       & 354     & 4  \\ \hline
4 & lodash-3.10.1       & 312    & 76 \\ \hline
5 & silverpeas-6.0.2      & 280  &1   \\ \hline
6 & Python-2.7.13       & 248    &141  \\ \hline
7 & Python-2.7.16       & 224    &117 \\ \hline
8 & jackson-databind-2.6.7.1 & 215  &13 \\ \hline
9 & jackson-databind-2.9.6      & 192  &12     \\ \hline
10 & Python-2.7.12      & 185    &107 \\ \hline
\noalign{\hrule height 1pt}
\end{tabular}
\caption{\label{tab:vuln_packages}The most vulnerable packages (based on critical severity level).}
\end{table}

\shorten{
\begin{figure}[h]
    \centering
    \includegraphics[width=1.0\linewidth]{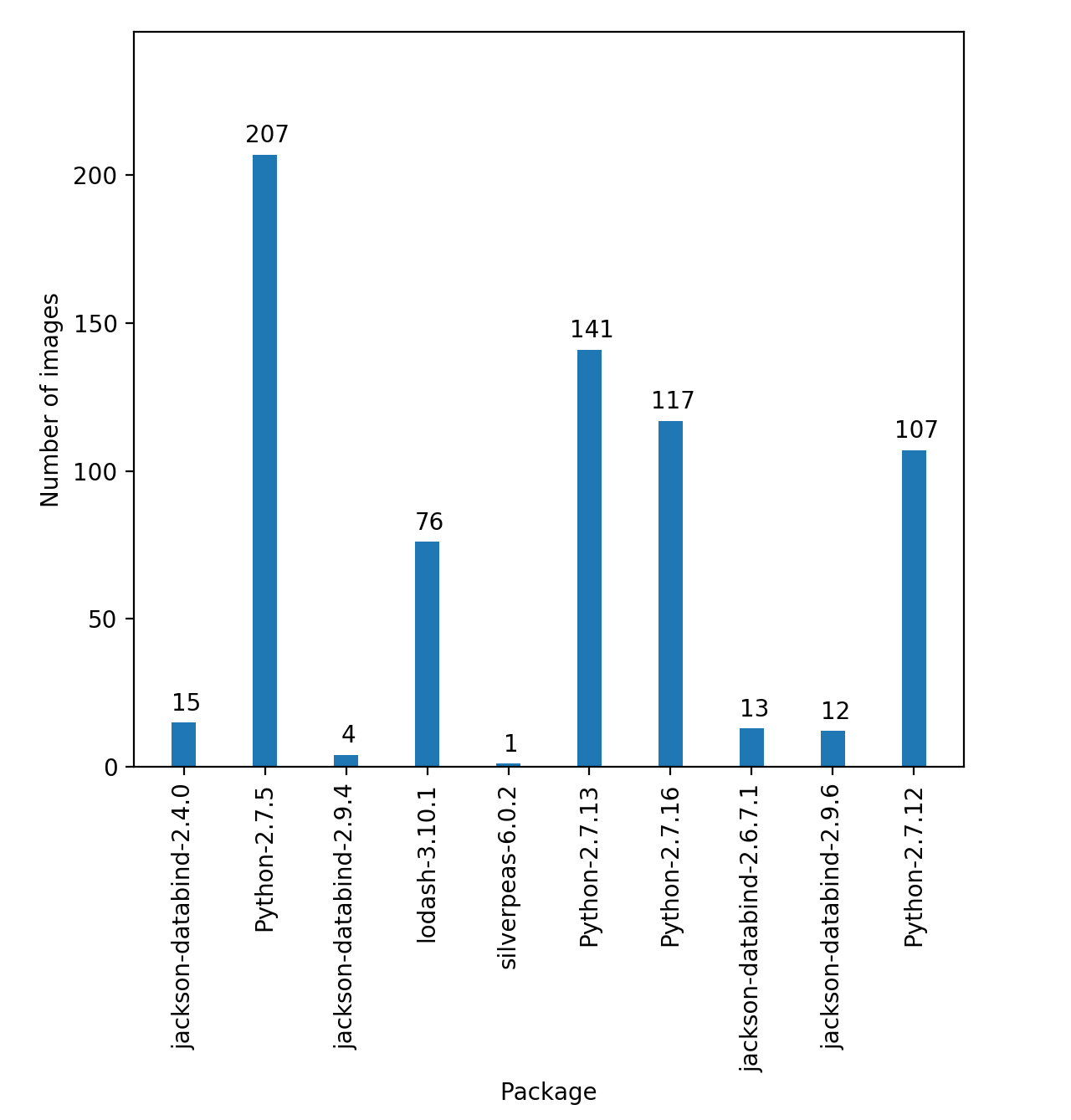}
    \caption{Number of images that uses the most vulnerable packages.}
    \label{fig:packages}
\end{figure}
}

\begin{table*}[]
\resizebox{\textwidth}{!}{%
\begin{tabular}{rlr|r|r|r|r|r|r}
\noalign{\hrule height 1pt}
  & \textbf{Package}  &   \textbf{Critical} & \textbf{High} & \textbf{Medium} & \textbf{Low} & \textbf{Negligible} & \textbf{Unknown} &  \parbox{1.5cm}{\centering \vspace{1.0mm} \textbf{Image count} \vspace{1.0mm}}\\ \noalign{\hrule height 1pt}
1 & tar-1.29b-1.1 & 0 & 0 & 0 & 0 & 482 & 0 & 241    \\ \hline
2 & coreutils-8.26-3 & 0 & 0 & 0 & 0 & 240 & 0 & 240        \\ \hline
3 & libpcre3-2:8.39-3 & 0 & 0 & 0 & 0 & 956 & 0 & 239        \\ \hline
4 &login-1:4.4-4.1 & 0  & 0 & 0 &0 & 714 &0 &238      \\ \hline
5 &passwd-1:4.4-4.1 &0 & 0 & 0 & 0 & 708 & 0 & 236      \\ \hline
6 &sensible-utils-0.0.9 & 0 & 0 & 103 & 0 & 0 & 111 &214       \\ \hline
7 &libgcrypt20-1.7.6-2+deb9u3 & 0 & 0 & 0 & 0 & 211 & 0 & 211      \\ \hline
8 &libgssapi-krb5-2-1.15-1+deb9u1 & 0 & 0 & 0 & 0 & 621 & 0 & 207    \\ \hline
9 &libk5crypto3-1.15-1+deb9u1 &0  & 0 & 0 & 0 & 621 & 0 & 207        \\ \hline
10 &libkrb5-3-1.15-1+deb9u1 & 0& 0 & 0 & 0 & 621 & 0 & 207  \\ \hline
\noalign{\hrule height 1pt}
\end{tabular}}
\caption{\label{tab:vuln_packages_all}Vulnerabilities in the most used packages.}
\end{table*}

From Table \ref{tab:vuln_packages}, one can observe that the Python packages are by far the most used packages, and therefore they expose the biggest impact regarding the threat landscape. The lodash-3.10.1 package is found in 76 images. This package contains the prototype pollution vulnerability affecting JavaScript code, which also is the most represented vulnerability in Table \ref{tab:vulns}. Further, the jackson-databind package is represented with four different versions in Table \ref{tab:vuln_packages} (entry 1, 3, 8 and 9). This package is used to transform JSON objects to Java objects (Lists, Numbers, Strings, Booleans, etc.), and vice versa. In total, these packages are used by 44 images: a relatively low amount compared to the usage of the Python packages. Finally, the silverpeas-6.0.2 package contains 280 critical vulnerabilities and is only used by a single image: the silverpeas image on Docker Hub.\footnote{\href{https://hub.docker.com/\_/silverpeas}{\textit{https://hub.docker.com/\_/silverpeas}}}

\subsection{Vulnerabilities in popular packages}\label{sec:vulns_in_pop_packages}
\vspace{-0.2cm}
When considering the packages that have the most critical vulnerabilities (Table \ref{tab:vuln_packages}), some of the packages are only used by a few images (like the silverpeas package). Therefore, Table \ref{tab:vuln_packages_all} is presented, as it is desirable to see what the vulnerability distribution is like in the most popular packages. The table shows the most used packages and the number of vulnerabilities that are present in them, considering all security levels. The image count column contain the number of images that use this package.

As observable from Table \ref{tab:vuln_packages_all}, the most used packages are not containing any critical, high, medium or low vulnerabilities (except for one entry). However, they are containing a vast number of negligible vulnerabilities, which is of less significance from a security point of view, as mentioned in previous sections.

\vspace{-0.2cm}
\section{Conclusions  and future work}
\vspace{-0.2cm}
This paper summarizes the findings that we reported in a longer and much more detailed work \cite{WistHelsem2020}. We studied the vulnerability landscape in Docker Hub images by analyzing 2500 Docker images of the four image repository categories: official, verified, certified images, and community. We found that as many as 82\% of certified images contain at least one high or critical vulnerability, and that they are the most vulnerable when considering the median value. Official images came out as the most secure image type with 45.9\% of them containing at least one critical or high rated vulnerability. Only 17.8\% of the images did not contain any vulnerabilities, and we found that the community images are the most exposed as 8 out of the top 10 most vulnerable images are community images. 

Concerning the technical specifics about the vulnerabilities, we found that the top five most severe vulnerabilities are coming from two of the most popular scripting languages, JavaScript and Python. Vulnerabilities in the Lodash library and vulnerabilities in Python packages are the most frequent and most severe. Furthermore, the vulnerabilities related to execution of code and overflow are the most frequently found critical vulnerabilities. Our scripts and tools are available from \cite{WistHelsem2020} and from the GitHub repository.
 
For the future work we first propose two improvements that are beyond our control, and are mostly connected with the maintenance of all 3.5 million images at the Docker Hub web site: 1. There is a need for a complete and well-documented endpoint for image data gathering;, and 2. There is a need for improvements on the Docker Hub web pages to make it possible to access all images through navigation.

Concerning improvements of this work, we consider a future analysis that will run over a more extended period. All previous studies conducted in this field, as well as ours, have only analyzed vulnerabilities in Docker Hub images captured from one single data gathering. Thus, changes in the data set over time are still not investigated. This type of analysis could reveal more in-depth details about the characteristics and evolution of the vulnerability landscape. 

Lastly, we suggest future work to be targeting the false positives and false negatives in container scanners by integrating machine learning into container scanners.

\bibliographystyle{IEEEtran}
\bibliography{main}

\shorten{

} 

\shorten{
\appendices

\section{Experimental data}
\lipsum[1-4]
\section{Theorem proofs}
\lipsum[5-6]

\vspace{12pt}
\color{red}
IEEE conference templates contain guidance text for composing and formatting conference papers. Please ensure that all template text is removed from your conference paper prior to submission to the conference. Failure to remove the template text from your paper may result in your paper not being published.

} 

\end{document}